\documentclass[9pt]{book}

\usepackage[dvips]{graphicx,color}
\usepackage{makeidx,universe}

\makeauthorindex

\BookTitle{The proceedings of the Physics of Accreting Compact Binaries}

\CopyRight{}

\begin{document} 

\pagenumbering{arabic}

\chapter{%
Seismology of Rapidly Rotating Accreting White Dwarfs}

\author{\raggedright \baselineskip=10pt%
{\bf Dean M. Townsley,$^{1}$
Phil Arras,$^{2}$
and 
Lars Bildsten$^{3}$}\\ 
{\small \it %
(1) Department of Physics and Astronomy, University of Alabama, Tuscaloosa,
AL 35487-0324, USA\\
(2) Department of Astronomy, University of Virginia, Charlottesville, VA
22904-4325, USA\\
(3) Department of Physics and Kavli Institute for Theoretical Physics,
University of California, Santa Barbara, CA 93106-9530, USA
}
}


\AuthorContents{Dean M. Townsley, Phil Arras, and Lars Bildsten } 

\AuthorIndex{Townsley}{D.M.} 

\AuthorIndex{Arras}{P.} 

\AuthorIndex{Bildsten}{L.} 

     \baselineskip=10pt
     \parindent=10pt

\section*{Abstract} 

A number of White Dwarfs (WDs) in cataclysmic binaries have shown brightness
variations consistent with non-radial oscillations as observed in isolated
WDs.
A few objects have been well-characterized with photometric campaigns in the hopes of gleaning information about the mass, spin, and possibly internal structural characteristics.
The novel aspect of this work is the possiblity to measure or constrain the interior structure and spin rate of WDs which have spent gigayears accreting material from their companion,
undergoing thousands of nova outbursts in the process.
In addition, variations in the surface temperature affect the site of mode driving, and provide unique and challenging tests for mode driving theories previously applied to isolated WD's.
Having undergone long-term accretion, these WDs are expected to have been spun up.
Spin periods in the range 60-100 seconds have been measured by other means for two objects, GW Lib and V455 And. Compared to typical mode frequencies, the spin frequency may be similar or higher, and the
Coriolis force can no longer be treated as a small perturbation on the fluid motions.
We present the results of a non-perturbative calculation
of the normal modes of these WDs, using interior thermal structures
appropriate to accreting systems.  This includes a discussion of the
surface brightness distributions, which are strongly modified from the
non-rotating case.
Using the measured spin period of approximately 100 seconds,
we show that the observed pulsations from GW Lib are consistent with the three lowest azimuthal order rotationally modified modes that have the highest frequency in the stellar frame.
The high frequencies are needed for the convective driving, but are then apparently shifted to lower frequencies by a combination of their pattern motion and the WD rotation.

\section{Introduction} 

There are now more than a dozen systems known in which the accreting WD
exhibits pulsational variations in its brightness consistent with non-radial oscillations as observed in isolated WDs \cite{dmt_WingetKepler2008}.
The first discovered and best studied example is GW Lib \cite{dmt_vanZetal00}.
We
performed an analysis of GW Lib under the assumption of slow rotation
\cite{dmt_TownArraBild04}, but pointed out at the time that these
WDs, having spent gigayears accreting matter via an accretion disk, are
likely to have been spun up.  The framework for treating spin in these cases
was developed in the context of ocean waves on neutron stars
\cite{dmt_Bildetal96}, and with certain caveats can be extended to uniformly
rotating WDs.  With the advent of a measured spin and mass for GW Lib
\cite{dmt_vanSpaandonketal10}, enough parameters are known about this object that we
can construct a consistent picture of the identity of its observed
pulsations.  A critical check is whether or not these appear consistent with
being driven by their interaction with the surface convection zone
\cite{dmt_ArraTownBild06}.

\section{White Dwarf Thermal State} 

A critical ingredient for modelling the pulsation due to global normal modes is the internal thermal state of the accreting WD, which is quite different than an isolated WD \cite{dmt_TownBild04}.
Whereas an isolated WD's surface temperature/flux, $T_{\rm eff}$, is directly
relate to its core temperature, $T_c$ (see e.g. \cite{dmt_Chabetal00}), in an
accreting system, $T_{\rm eff}$ is determined principally by the mass
transfer rate onto the WD and its overall mass, $M_{\rm WD}$
\cite{dmt_TownsleyGaensicke09}.  Thus measurement of $T_{\rm eff}$ provides only
weak constraints on $T_c$, and a thermal structure of even the WD envelope
from isolated systems is entirely inappropriate.  Accreting WDs with $T_{\rm
eff}$ near the instability strip have significantly lower $T_c$ than isolated
WDs at similar $T_{\rm eff}$.

The WD is ``re-heated'' by the accretion up to an equilibrium $T_c$ determined
by the balance of heat input by accretion from compression of the accreted
material (\emph{not} from infall energy) and a low level of nuclear fusion
leading up to the nova runaway, and loss of energy via observed radiation
between nova outbursts \cite{dmt_TownBild04}.  The speed that the
re-heating takes place is also dependent on the accretion rate, being faster
for higher $\langle\dot M\rangle$.  Above the period gap, where $\langle\dot
M\rangle\sim
10^{-9}M_\odot$~yr$^{-1}$, there is not enough time for the WD to reach
equilibrium.  However, below the period gap, the $\langle\dot M\rangle\sim
10^{-11}M_\odot$~yr$^{-1}$ is nearly constant for approximately 4 Gyrs
\cite{dmt_Howeetal01}, which is more than enough time for the WD to come to
equilibrium, even at such a low $\langle\dot M\rangle$.

\begin{figure}[t]
\centerline{\includegraphics[height=6cm]{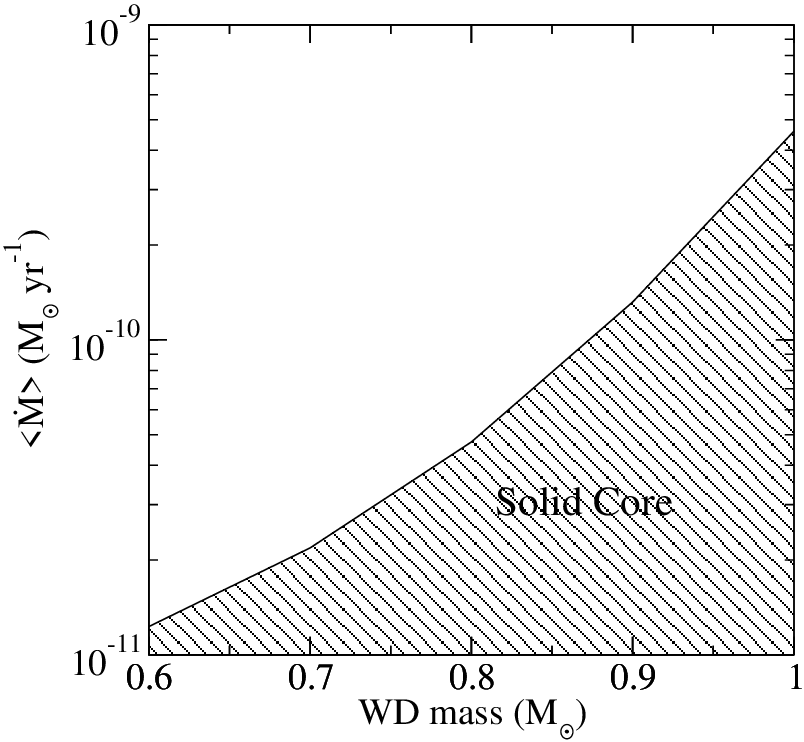}}
\vspace{-1pc}
\caption{\label{dmtfig:solid_core} Region of parameter space in average
accretion rate, $\langle \dot M\rangle$, and WD mass in which the center of
the WD is expected to be solid (crystalline).  This is determined by the
equilibrium central temperature.}
\end{figure}

As an example of the critical information garnered from the thermal WD state
is shown in Figure \ref{dmtfig:solid_core}, where the presence or absence of
a solid core is indicated as a function of the $M_{\rm WD}$ and $\langle\dot
M\rangle$.  Here we assume a carbon-oxygen core.
Note that, given the typical $\langle\dot M\rangle$ of a few $\times
10^{-11}M_\odot$~yr$^{-1}$, accreting pulsating WDs have solid cores at much
lower masses than their isolated pulsating counterparts \cite{dmt_MontgomeryWinget1999}.  This is due to the
surface luminosity being provided by compression of surface material due to
accretion rather than cooling of the core, leading to a low $T_c$ as
mentioned above.

\section{Seismology Under Rapid Rotation and Driving}

When the spin frequency becomes similar or larger than the mode frequency,
rotation cannot be treated in a perturbative fashion.  The spin breaks
the spherical symmetry of the problem and introduces a Coriolis force into
the equations for fluid motion.  Methods for treating this case were adapted from
Geophysics for Neutron star oceans by \cite{dmt_Bildetal96}, and we
will make use of this method here.  The approximation is called the
``traditional approximation'' and depends upon the assumption that $N^2\gg
\Omega \omega k_r/k_t$ and $k_r\gg k_t$ where $N$ is the Brunt-Vaisala
(buoyancy) frequency, $\Omega$ is the stellar spin frequency, $\omega$ is the
mode frequency, and $k_r$ and $k_t$ are the radial and tangential wave
numbers respectively.  These conditions are satisfied in the WD except near
the center, where $N$ is small.  However, for the models studied here, the
center is solid, so that waves do not propagate there.

The traditional approximation results in a separation of the angular from the radial equations for the perturbed fluid motion equations.
The differential equation for the angular part, which previously had spherical harmonic solutions, is replaced by Laplace’s tidal equation \cite{dmt_Bildetal96}.
Solutions to this equation depend on the parameter $q=2\Omega/\omega$ in
addition to $\ell$ and $m$.  The radial differential equation for the mode
shape (namely giving $\xi_r(r)$ and $\xi_t(r)$, the radial and tangential
motions) recovers a form just like the non-rotating case except that
$\ell(\ell+1)$ is replaced by an eigenvalue $\lambda$ determined by Laplace's
tidal equation and depending on $q$, $\ell$ and $m$.

\begin{figure}[t]
\centerline{\includegraphics[height=6cm]{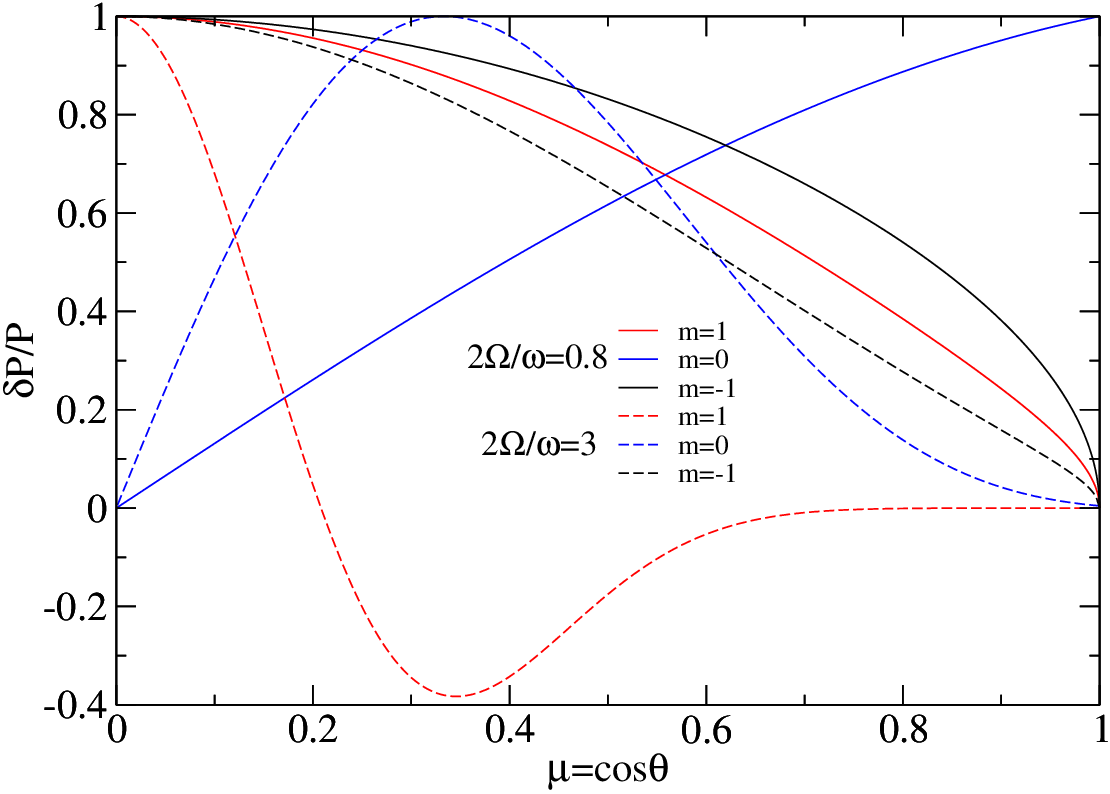}}
\vspace{-1pc}
\caption{\label{dmtfig:surf_efunc}
Rotationally modified surface eigenfunctions and their dependence on the
ration of spin to mode frequency $q=2\Omega/\omega$.
}
\end{figure}

The action of the Coriolis force is to tend to ``squeeze'' modes towards the
equator.  This can be seen from Figure \ref{dmtfig:surf_efunc}, which shows
the form of several of the angular eigenfunctions in $\mu=\cos\theta$ for two different values of $q$,
where $\theta$ is the classical polar coordinate so that $\mu=0$ is the equator ond $\mu=1$ is the pole.
This
function also determines the visibility of perturbations on the surface, so
that the strongest perturbations for the $m=1$ modes, for example, will be
concentrated near the equator and this will become more so at higher $q$.
Higher $q$ corresponds to either higher $\Omega$ or lower $\omega$ (at fixed
$\Omega$) and therefore higher radial order g-modes.

\begin{figure}[t]
\centerline{\includegraphics[height=6cm]{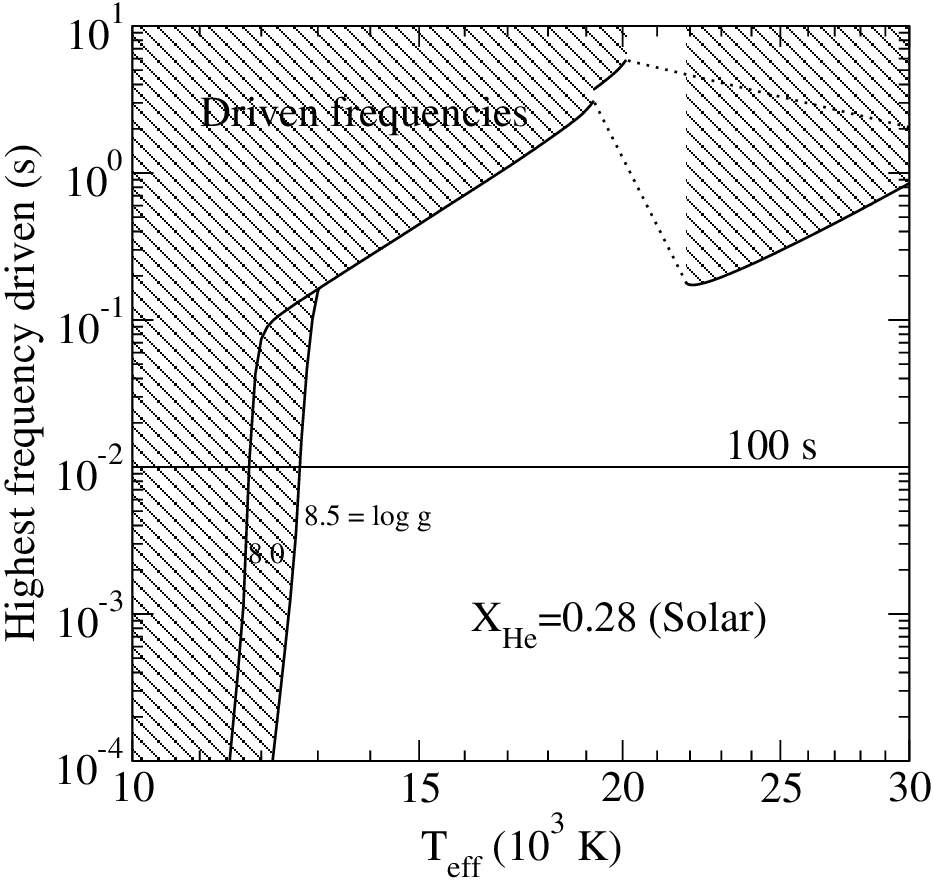}}
\vspace{-1pc}
\caption{\label{dmtfig:driving}
Diagram showing frequencies of driven modes (shaded region) for a range of WD
effective temperatures.
}
\end{figure}

On WDs, g-mode oscillations are known to be driven by the interaction of the
modes with the surface convection zone \cite{dmt_GoldWu99}.  This feedback
process drives modes whose frequency is greater than approximately the
thermal time at the base of the convection zone.  Since the highest frequency
g-modes in a WD have frequencies of around 0.005~Hz or so, they are not
excited above $T_{\rm eff}\approx 12-13$~kK.  There are also important
effects of both the abundance of the accreted material (and therefore the
convection zone) and the WD mass \cite{dmt_ArraTownBild06}.  Figure
\ref{dmtfig:driving} shows roughly the frequencies expected to be driven for
various $T_{\rm eff}$ values.  Also indicated is frequency corresponding to a
mode with a 100 second period.

\section{Application to GW Lib} 

\begin{figure}[t]

\begin{center}
\includegraphics[width=.49\textwidth]{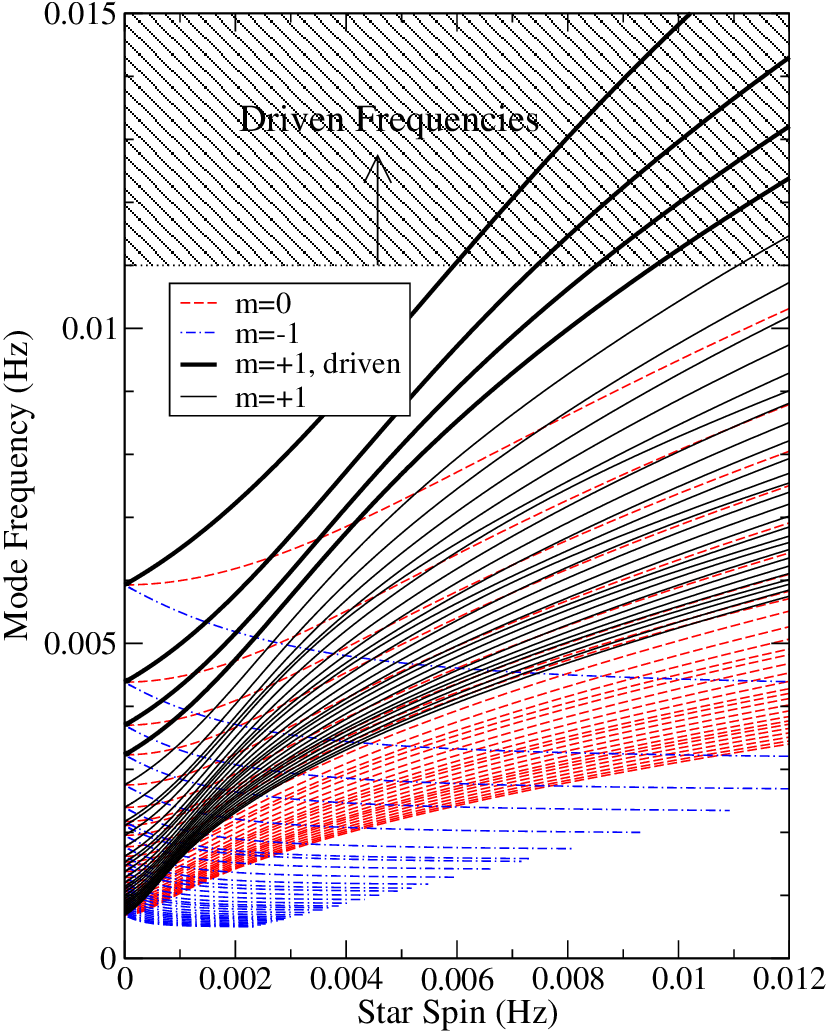}%
\includegraphics[width=.49\textwidth]{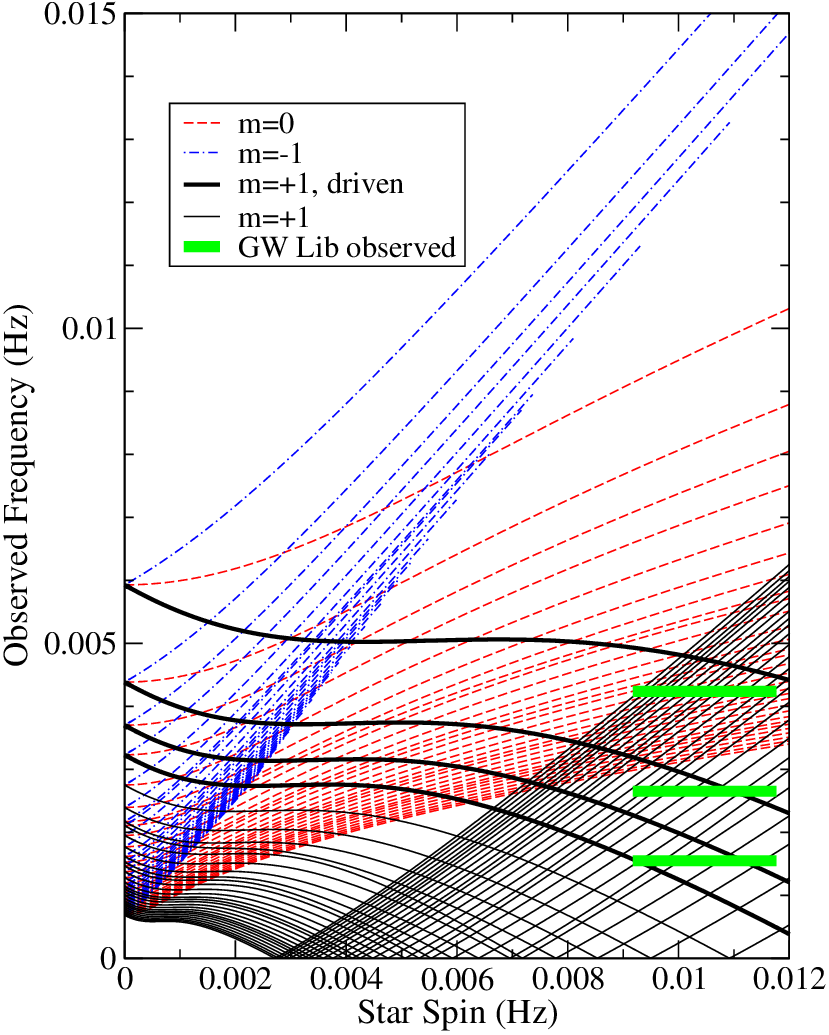}
\end{center}
\caption{\label{dmtfig:modefreq}
      Rotationally modified g-mode frequencies as a function of spin for
$\ell=1$.  The left panel shows the frequencies in the (rotating) star's frame
and the right panel shows the frequency observed by a
stationary (non-rotating) observer.  Also indicated in the star frame is a driving range
which would excite only the highest 4 modes at a stellar spin of 0.01~Hz.
These modes are highlighted in both figures.  Finally the frequencies
observed in GW Lib are indicated with the observed spin frequency and its
uncertainty.
      }
\end{figure}

In the case of GW Lib, \cite{dmt_vanSpaandonketal10} measured the spin period from
the width of absorption lines to be approximately 100~s.  This implies that
the g-modes will be strongly modified by rotation.  The $T_{\rm eff}$ of GW
Lib has been measured \cite{dmt_SzkoGWLib02} from UV spectroscopy to be 13-16~kK,
but is subject to some ambiguity due to a single-temperature fit not
reproducing certain features of the spectrum.  We carried out a calculation
of the rotationally-modified g-modes for an object similar to our previous
study of GW Lib \cite{dmt_TownArraBild04} with $M=0.9M_\odot$ and an accreted
layer mass of half the nova ignition mass.  The mode frequencies in the
star's frame (the fluid frame) for $\ell=1$ are shown in the left panel of
Figure \ref{dmtfig:modefreq} as a function of the stellar spin.  The modes
are split as expected according to their $m$ values at low spin, but then
become strongly shifted at higher spins, overlapping modes of other $m$
values.  This features makes identifying mode structure with no knowledge of
the possible WD spin challenging.

Also indicated in the left panel is a possible limit on the driven mode
frequencies which would cause only the 4 $m=1$ g-modes with the highest
frequencies to be driven for $f_{\rm spin}=0.01$~Hz.  Modes with higher
$\ell$ values may also be driven, but, as in isolated WDs, would not show up
as strongly as surface perturbations due to cancellation.  These modes
frequencies are highlighted with heavy lines in both the left and right
panels of Figure \ref{dmtfig:modefreq}.

Both the $m=-1$ and $m=+1$ modes form a brightness pattern which propagates
with respect to the surface of the WD, in pro- and retrograde directions
respectively.  When combined with the spin, this causes a fixed observer to
see a different frequency of oscillation than that of the oscillation on the
WD.  The observed frequency is $f_{\rm obs} = |f_{\rm mode}-mf_{\rm star}|$.
The $m=0$ mode has the same frequency on the star and to an observer because
it has no rotating pattern.  Taking the $m=1$ modes as an example, at low
spin the retrograde propagation of the modes with $\omega \gg \Omega$ causes
the observed frequency to be near the non-rotating g-mode frequency, but
slightly lower.  As the spin becomes relatively stronger, the mode is
``dragged'' along with the star, until it eventually passes through being
stationary (zero frequency) and becomes prograde as seen by the observer.

The three thick horizontal lines in the bottom right of the right panel of
Figure \ref{dmtfig:modefreq} indicate the observed frequencies of pulsations
on GW Lib.  The horizontal extent of these lines corresponds to the
uncertainty in the spin given by \cite{dmt_vanSpaandonketal10}.  Note that the present
WD model has \emph{not} been fit to these lines.  The relative spacing among
these lines is sensitive mainly to a combination of the $\langle\dot M\rangle$, the accreted layer
mass, and $M_{\rm WD}$.  The near consistency without a fit indicates that
there is some combination of these that will allow the simultaneous fit of
these periods at a spin frequency that is consistent with the observations.
It is also consistent with the mode driving that only these three modes are
driven, since they are the highest frequency modes in the star frame.
Additionally, higher radial order modes, which have a smaller $\omega$, will
be more tightly constrained to the equator, possibly limiting their
visibility.

\section{Conclusion} 

Motivated by the recent measurement of the spin of the accreting WD GW Lib,
we have presented a consistent picture for the pulsations observed.  In this
picture the 3 independent pulsations observed are due to the 3 highest
frequency $\ell=1$ rotationally modified g-modes.  These modes have $m=1$ and
have a retrograde pattern propagation so that to a fixed observer they have
an apparently lower frequency than on the star.  The higher frequency of
these modes in the star's frame (above 0.01~Hz) also allows GW Lib to enter
the instability strip at higher $T_{\rm eff}$ than it would have if it were
not rotating rapidly.

While here we have shown the structure of $l=1$ g-modes for a rapidly
rotating object with similar properties to those expected of GW Lib, we have
only shown that a consistent fit is plausible.  It is expected that an
actual fit of the observed mode frequencies, along with the observed
constraints on spin, $T_{\rm eff}$ and mass will provide a measurement of the
mass of the accreted layer of solar abundance material.  The presence of
additional modes is still possible, and these may be at lower or similar
frequency to the lowest frequency already observed, as the next few radial
order modes are expected to switch from retrograde to prograde from the
observer's point of view.



%
%
%
%
%
%

\end{document}